\documentstyle[12pt]{article}
\newcommand{\nn}{\nonumber \\}
\renewcommand{\o}{\over}
\newcommand{\ri}{\right}
\newcommand{\lf}{\left}
\newcommand{\Bl}{\biggl}
\newcommand{\Br}{\biggr}
\newcommand{\bl}{\Bigl}
\newcommand{\br}{\Bigr}

\newcommand{\del}{\partial}
\newcommand{\cob}{\delta}    
\newcommand{\al}{\alpha}

\newcommand{\bt}{\beta}

\newcommand{\th}{\theta}
\newcommand{\Ga}{\Gamma}

\newcommand{\si}{\sigma}

\newcommand{\om}{\omega}

\newcommand{\riya}{\rightarrow}

\newcommand{\Tr}{{\rm Tr}}

\newcommand{\ket}{\langle}
\newcommand{\bra}{\rangle}

\renewcommand{\dag}[1]{#1^\dagger}

\renewcommand{\^}{\hat}
\newcommand{\half}{{1 \over 2}}

\renewcommand{\t}[1]{\tilde{#1}}

\newcommand{\tQ}{\tilde{Q}}
\newcommand{\Z}{{\bf Z}}
\newcommand{\T}{{\bf T}}
\newcommand{\vX}{\vec{X}}
\newcommand{\M}{{\cal M}}

\newcommand{\df}{\stackrel{\rm def}{=}}

\newcommand{\msc}[1]{\mbox{\scriptsize #1}}
\newcommand{\dsp}{\displaystyle}

\newcommand{\e}{\mbox{{\bf e}}}
\newcommand{\va}{\mbox{{\bf a}}}
\newcommand{\bc}{\mbox{{\bf C}}}
\newcommand{\bz}{\mbox{{\bf Z}}}

\newcommand {\eqn}[1]{(\ref{#1})}

\newcommand{\be}{\begin{equation}}\newcommand{\ee}{\end{equation}}
\newcommand{\bea}{\begin{eqnarray}} \newcommand{\eea}{\end{eqnarray}}
\newcommand{\ba}[1]{\begin{array}{#1}} \newcommand{\ea}{\end{array}}

\newcommand{\cleqn}{\setcounter{equation}{0}}
\makeatletter
\@addtoreset{equation}{section}

\makeatother

\begin{document}
\vskip 7mm
\renewcommand{\thefootnote}{\fnsymbol{footnote}}
\font\csc=cmcsc10 scaled\magstep1
{\baselineskip=14pt
 \rightline{
 \vbox{\hbox{hep-th/9806001}
       \hbox{UT-820}
       }}}

\vfill
\baselineskip=20pt
\begin{center}
\centerline{\Huge Fractional Strings in $(p,q)$ 5-brane} 
\vskip 2.5mm
\centerline{\Huge   and}
\vskip 3mm
\centerline{\Huge   Quiver Matrix String Theory}
\vskip .8 truecm

Kazumi Okuyama and Yuji Sugawara\\
{\sf okuyama@hep-th.phys.s.u-tokyo.ac.jp~,~
sugawara@hep-th.phys.s.u-tokyo.ac.jp}

\vskip .6 truecm
{\baselineskip=15pt
 {\it Department of Physics,  Faculty of Science\\
  University of Tokyo\\
  Bunkyo-ku, Hongo 7-3-1, Tokyo 113-0033, Japan}
}
\vskip .4 truecm

\end{center}

\vfill
\vskip 0.5 truecm

\begin{abstract}
\baselineskip 6.7mm
We study the $(p,q)$5-brane dynamics from the viewpoint of 
Matrix string theory 
in the T-dualized ALE background.
The most remarkable feature  in the $(p,q)$5-brane
is the existence of  ``fractional string'', 
which appears as the instanton of 
5-brane gauge theory. 
We approach to the physical aspects of fractional string 
by means of the two types of Matrix string probes: One of which is that  
given in \cite{Witten}. As the second probe we present  
the Matrix string theory describing the fractional string itself.  
We calculate the moduli space metrics 
in the respective cases and argue on  the specific behaviors 
of fractional string. 
Especially, we show that the ``joining'' process of fractional strings 
can be realized as the transition from the Coulomb branch to the Higgs 
branch of the fractional string probe.
In this argument,
we emphasize the importance of some monodromies related with 
the $\theta$-angle of the 5-brane gauge theory. 

\end{abstract}

\setcounter{footnote}{0}
\renewcommand{\thefootnote}{\arabic{footnote}}

\newpage
\baselineskip 7mm

\section{Introduction}
\cleqn
\hspace*{4.5mm}
Analyses of the brane dynamics have been playing 
a central role in the studies on string duality and M-theory.
Among others,  the studies  of 5-brane have  a primary importance and
gives rich products. 
This is because the 5-brane 
is the magnetic dual of string (and also the dual of the
M-theory membrane), and from the viewpoint of Matrix string theory \cite{DVV},
the degrees of freedom of 5-brane are  naturally incorporated into the theory
by considering some matter fields (hypermultiplets) \cite{NS5A,diacone}.

In the limit when the gravitational interaction to the bulk theory 
decouples, the 5-brane dynamics leads  to the 6-dimensional 
``new'' quantum theory \cite{new-th}. The most remarkable feature 
of this 6-dimensional theory is the existence of a non-local excitation 
(non-critical string). In the set up from the IIB 5-branes, 
whose low energy effective theory is a 6-dimensional 
gauge theory, this non-critical string appears as the instanton.
In the system of parallel NS5-branes (D5-branes)  with the vanishing 
IIB $\theta$-angle, this naturally identified with the fundamental 
string (D-string) trapped inside their world volumes \cite{new-th}.
However, as is emphasized in \cite{kol2}, in the $(p,q)$ 5-brane\footnote
     {We use the term ``$(p,q)$ 5-brane'' as the meaning of the bound 
state of $p$ D5s and $q$ NS5s, according to the convention in \cite{Witten}.} 
cases (or equivalently, NS5s or D5s with some rational $\theta$-angle)
we face   different circumstances.  One might imagine
the instanton string can be constructed  by the simple
$SL(2,\bz)$-duality transformation 
from the fundamental string (in the NS5 theory, and, of course,
D-string for D5). But this is {\em not\/} correct. 
It is known  that
the instanton string in the $(p,q)$ 5-brane  has the following 
tension \cite{kol2,kol}
\begin{equation}
T^{{\rm instanton}}_{p,q}={{\rm Im}\tau\o |a+b\tau|}T ,
~~~ ((p,q)= r(a,b), ~~ 
\mbox{$a,b$ : coprime})
\end{equation}
where $T$ denotes the tension of the fundamental string and $\tau$ is the IIB
complex coupling 
$\dsp \tau = \frac{\theta_{\msc{IIB}}}{2\pi}+ i \frac{1}{g_s} $.
For generic values of $\tau$, 
this does not coincide with {\em any\/} string tension 
of the $SL(2,\bz)$-multiplet of BPS strings 
(so-called ``$(r,s)$-string''): $T_{r,s}=|r-s\tau| T$. 
Especially, under  the decoupling limit $\tau \longrightarrow i\infty$, 
we have $T^{{\rm instanton}}_{p,q} = T/b$, which is $b$-times smaller
than that of the fundamental string. 
So it is plausible to call it as the ``fractional string''.

We intend in this paper to study the property of fractional 
string from the stand point of Matrix theory \cite{BFSS}.
To this aim, it is a natural set up  to take the fractional string as 
the Matrix string probe.    
But this is not so easy, because, as we just commented,  
the fractional string is not a D-brane (nor the object obtained from
a D-brane  by U-duality) in the usual sense. 
Therefore  we shall take the T-dualized framework - 
M-theory compactified on some ``twisted'' orbifold \cite{Witten}, 
and use the technique of  the quiver Matrix theory introduced 
in \cite{d-moore,douglas}. 

In section 2 we shall begin by reviewing the M-theory picture corresponding
to the  $(p,q)$ 5-brane and its Matrix theory realization given in
\cite{Witten}. We calculate the moduli space metric and discuss how we
can observe the exciteations of fractional string in this framework.

In section 3 we construct the quiver Matrix theory defined on the fractional
string probe. This is a more direct approach to the dynamics of 
this object  than that of \cite{Witten}. We will show 
some  existence of  monodromies  related 
with the $\theta$-angle of the 6-dimensional gauge theory,
which will play an important role in our analysis of the moduli space.
We find out a different  structure in  its moduli space
from that of the usual D-brane probe. 
This observation of ours
will clarify  the peculiar behaviors of fractional string.

Section 4 is devoted to the discussions and some comments.

~

\section{Witten's Matrix String Probing $(p,q)$ 5-brane}
\cleqn

\subsection{M-theory Picture of the $(p,q)$ 5-brane}
\hspace*{4.5mm}
We shall begin from  the T-dualized  construction of $(p,q)$
5-brane theory introduced in \cite{Witten}.
The claim is that,
under the T-duality along one of the transversal direction,
$(p,q)$ 5-brane in IIB string theory is described 
by M-theory compactified on the ``twisted'' orbifold   
\be
X_{p,q} \df (\bc^2 \times S^1)/\bz_q ,
\ee
where the $\bz_q$-action is defined by 
\be
\left\{
\begin{array}{lll}
z_1 & \longrightarrow & \om z_1 \\
z_2 & \longrightarrow & \om^{-1} z_2 \\
u & \longrightarrow & \om^{-p} u
\end{array}
\right. ~~~ (\om \df e^{2\pi i/q }) .
\label{Zq}
\ee
$(z_1,z_2)$ is the coordinate of $\bc^2$ and $u$ denotes that of $S^1$.  
Let us assume that $\gcd (p,q) =r$, and set $(p,q) =r(a,b)$.
In this case $X_{p,q}$ has an $A_{r-1}$-singularity, which corresponds 
to the $U(r)$-gauge symmetry on the 5-brane. 

Roughly speaking, the above  statement can be explained as follows:
Consider the NS5 in IIB string theory wrapping around 
the $0,\ldots,5$-th directions.  
If we take the T-duality along one of the transversal direction, 
say, the 6-th direction, we obtain IIA string theory compactified on 
the Taub-NUT space with the $S^1$-fibration along the dual  
6-direction (we shall call it the  ``TN-direction'');
\be
\mbox{NS5} ~ \stackrel{T^6}{\longleftrightarrow} ~ 
IIA/TN \cong M/(TN \times S^1) ~~~\mbox{(The TN-direction is the 6-th axis.)} 
\ee
Similarly, D5 is mapped  to the D6 by T-duality, and D6 is interpreted as 
M/TN with the TN-direction is the 11-th axis;
\be
\mbox{D5} ~ \stackrel{T^6}{\longleftrightarrow} ~ 
 D6 \cong M/(TN \times S^1) ~~~\mbox{(The TN-direction is the 11-th axis.)} 
\ee
In this way, we can expect  the bound state of $p$ D5s and $q$ NS5s 
corresponds to the $M/(TN \times S^1)$ with the TN-direction 
$a \e_{11} + b \e_{6}$.
 ($S^1$ is also wrapping around $c \e_{11} + d \e_{6}$, where
$c$, $d$ is the integers uniquely determined from $a$, $b$, so that
$\dsp \left(
\begin{array}{ll}
 a & c \\
 b & d
\end{array}
\right) \in SL(2,\bz)$.)
$X_{p,q}$ roughly has this $TN \times S^1$ structure
(under the limit when the TN-circle decompactifies)  
away from the singular point ($z_1= z_2=0$).

However, the role  of singularity is essential for our discussion.
At the singular point, the TN-circle shrinks (the vanishing cycle)
and we encounter  an  ``exotic''  circle  which has a fractional
radius \cite{Witten}. This is because,  by the identification
of the $\bz_q$-action \eqn{Zq}, the points $(z_1, z_2 ,u)=(0,0,u) $ 
are identified with $(0,0, \om^{-p} u) $, which produces 
a fractional circle. It is in fact  the origin of fractional string. 
Namely, the M2-brane wrapping arround this fractional circle 
appears as a   string possessing  the fractional tension.


\subsection{Witten's Matrix String and Screwing Procedure}
\hspace*{4.5mm}
Considering the compactification over another $S^1$ 
(say, the 5-th direction) : $M/(X_{p,q}\times S^1)$, 
one can introduce the DVV's Matrix string theory \cite{DVV}
which has the KK momentum along this extra $S^1$.
By means of  the U-duality ($5 \leftrightarrow 11$-flip), 
we can reduce this system to the D0-probes in $IIA/X_{p,q}$.
Now, the $\T^2$-fiber of $X_{p,q}$ is wrapping around 
the 5,6-th directions.
(TN-direction becomes $a \e_5 + b\e_6$.) 
It is the starting  point 
of the Matrix theory realization in \cite{Witten}.

Let us trace the construction of  this Matrix theory.
The bosonic part of the action of D0-branes in flat space
is 
\begin{equation}
S=T_{D0}\int dt ~\Tr\lf(\half \sum_{I=1}^{9}(D_0X^I)^2
+{T^2\o 4}\sum_{I,J=1}^{9}[X^I,X^J]^2  \ri)
\label{D0action}
\end{equation} 
where $T_{D0}=1/g_s^Al_s$ is the mass of D0-brane and $T=1/2\pi l_s^2$ is the
tension of the fundamental string.
We decompose nine-dimensional transverse space into ${\bf R}^4\times
{\bf R}\times {\bf C}^2$,
\be
\ba{l}
\left\{
\begin{array}{lll}
&{\bf R}^4&:  X^i~~(i=1,2,3,4)\nn
&{\bf R}  &:  X^5 \nn
&{\bf C}^2&:  X^a~~(a=6,7,8,9) \nn
\end{array}
\right.   \\
~~~~~Q=X^6+iX^7,~\tQ=X^8+iX^9 
\ea
\ee
First we remark that
$X_{p,q}\equiv (\bc^2 \times S^1)/\bz_q 
 \cong ( {\bf R} \times  {\bf C}^2)/\Ga$, 
where the action of the Abelian group $\Ga=\{\al,\bt\}$ on 
${\bf R}\times {\bf C}^2$ is given by,
\begin{equation}
\begin{array}{llll}
\al: & Q\riya \om Q,&\tQ\riya \om^{-1}\tQ,&
 X^5\riya X^5-2\pi R_5p/q\\
\bt :& Q\riya Q,&\tQ\riya\tQ,&X^5\riya X^5+2\pi R_5 .
\end{array}
\end{equation}
So, our task is to construct the Matrix theory with the $\Ga$-invariance 
imposed.
The Chan-Paton factor is labeled by
the element of $\Ga$, and matrix element is expressed as 
$\ket g|X^I|g'\bra ~(g,g'\in \Ga)$. 
We can  represent an element of $\Ga$ in terms of the generator,
as $\al^s\bt^m=(s,m)$, where $s\in {\bf Z}_q$ and $m\in {\bf Z}$.
The transformation law of $X^I$ under $\Ga$ is expressed as 
\begin{eqnarray} 
&&\ket s+t,m+n|X^i|s'+t,m'+n\bra =\ket s,m|X^i|s',m'\bra \nn
&&\ket s+t,m+n|X^5|s'+t,m'+n\bra =\ket s,m|X^5|s',m'\bra +
  2\pi R_5(n-{p\o q}t) \ket s,m|s',m'\bra\nn
&&\ket s+t,m+n|Q|s'+t,m'+n\bra =\om^t\ket s,m|Q|s',m'\bra \nn
&&\ket s+t,m+n|\tQ|s'+t,m'+n\bra =\om^{-t}\ket s,m|\tQ|s',m'\bra 
\label{red}
\end{eqnarray}
where $\ket s,m|s',m'\bra =\cob_{s,s'}\cob_{m,m'}$.
By these relations \eqn{red},
we can define the ``reduced matrix element'' $\ket g|X\bra$ of matrix $X$ by
\begin{equation}
\ket s,m|X\bra := \ket s,m|X|0,0\bra ,
\end{equation}
in the same way as \cite{taylor}.
It is useful to make further 
the Fourier transformation with respect to $\Ga$.
Namely, one can transfer the basis of Chan-Paton Hilbert space from that for
$\Ga$ to that for the irreducible representations
of $\Ga$ (we denote it as $\Ga^*$),
which is labeled by $R=(k,\th) \in \bz_q \times \tilde{S}^1$. 
($\tilde{S}^1$ is the dual circle along the 5-th direction.)
The transformation coefficient $\ket R|g\bra$ $(g\in \Ga, R \in \Ga^*)$
is nothing but the chracter of the irrep. $R$:
\begin{equation}
\ket R|g\bra=\ket k,\th|s,m\bra=\om^{ks}e^{i\th(m-{p\o q}s)} ,
\end{equation} 
and one can immediately notice the following properties:
\begin{eqnarray}
&&\ket k,\th|s+q,m+p\bra=\ket k,\th|s,m\bra \nn
&&\ket k,\th+2\pi|s,m\bra=\ket k-p,\th|s,m\bra  .
\label{period}
\end{eqnarray}
Now, we can write down the matrix element 
with respect to the $\Ga^*$-basis: 
\begin{eqnarray}
&&\ket k,\th|X^i|k',\th'\bra= 2\pi q \cob_{k,k'}\cob(\th-\th') X^i_k(\th) \nn
&&\ket k,\th|X^5|k',\th'\bra= \left( -2\pi R_5i{d\o d\th}+ X^5_k (\th) \right)
    2\pi q \cob_{k,k'}\cob(\th-\th')  \nn
&&\ket k,\th|Q|k',\th'\bra= 2\pi q \cob_{k,k'-1}\cob(\th-\th') 
      Q_{k,k+1}(\th)\nn
&&\ket k,\th|\tQ|k',\th'\bra= 2\pi q \cob_{k,k'+1}\cob(\th-\th')
       \tilde{Q}_{k,k-1}(\th)
\end{eqnarray}
where we have introduced the matrix variables expressing 
the reduced matrix element for $\Ga^*$-basis;
\begin{eqnarray}
&&  X_k^i(\th) =\ket k,\th|X^i\bra\nn
&&  X_k^5(\th)=\ket k,\th|X^5\bra \nn
&&  Q_{k,k+1}(\th)=\ket k,\th|Q\bra \nn
&&  \tQ_{k,k-1}(\th)=\ket k,\th|\tQ\bra . 
\end{eqnarray} 
From (\ref{period}), these variables should satisfy 
the following monodromy (so-called the ``clock-shift''); 
\begin{eqnarray}
X_k^i(\th+2\pi)&=& X_{k-p}^i(\th) \nn
X_k^5(\th+2\pi)&=& X_{k-p}^5(\th)  \nn
Q_{k,k+1}(\th+2\pi) &=& Q_{k-p,k+1-p}(\th) \nn
\tQ_{k,k-1}(\th+2\pi) &=& \tQ_{k-p,k-1-p}(\th)
\label{pstep}
\end{eqnarray} 
Rescaling further $A_1=TX^5$ and
\begin{equation}
x^1=\t{R}_5\th={l_s^2\o R_5}\th , 
\end{equation}
we obtain the expression of the original D0-brane action 
written in the Fourier transformed variables:
\begin{equation}
S=T_{D1}{1\o q}\sum_{k=0}^{q-1}\int dt\int_0^{2\pi\t{R}_5}dx^1 \Tr_N L_k
\label{Zq-action}
\end{equation}
where $T_{D1}=T/g_s^B=TR_5/g_s^Al_s$ is the tension of D1-brane 
and $\Tr_N$ is the trace over $U(N)$ indices.
$L_k$ is given by
\begin{eqnarray}
L_k&=&-{1\o 4T^2}F_{\mu\nu,k}F^{\mu\nu}_k-{1\o 2}\Bl\{( D_{\mu}X^i_k)^2
 +|D_{\mu}Q_{k,k+1}|^2+|D_{\mu}\tQ_{k,k-1}|^2\Br\} \nn
&&-{T^2\o 2}\lf\{-\half [X_k^i,X^j_k]^2
+\bl|[X^i,Q]_{k,k+1}\br|^2+\bl|[X^i,\tQ]_{k,k-1}\br|^2
 +\bl|\mu_k^{{\bf C}}\br|^2+\lf(\mu_k^{{\bf R}}\ri)^2\ri\} .
\end{eqnarray}
Here  we have  defined
\begin{eqnarray}
D_{\mu}Q_{k,k+1}&=&\del_{\mu}Q_{k,k+1}+i(A_{\mu,k}Q_{k,k+1}
     -Q_{k,k+1}A_{\mu,k+1}) \nn
~[X^i,Q]_{k,k+1}&=&X^i_kQ_{k,k+1}-Q_{k,k+1}X^i_{k+1} \nn
\mu_k^{{\bf C}}&=&[Q,\tQ]_k \nn
  &=&Q_{k,k+1}\tQ_{k+1,k}-\tQ_{k,k-1}Q_{k-1,k} \nn
2\mu_k^{{\bf R}}&=&[Q,\dag{Q}]_k+[\tQ,\dag{\tQ}]_k \nn
 &=& Q_{k,k+1}\dag{Q}_{k,k+1}-\dag{Q}_{k-1,k}Q_{k-1,k}
    +\tQ_{k,k-1}\dag{\tQ}_{k,k-1}-\dag{\tQ}_{k+1,k}\tQ_{k+1,k} .
\end{eqnarray}
The action (\ref{Zq-action}) 
gives the well-known form of the  $U(N)$ $A_{q-1}$ quiver gauge theory 
on ${\bf R}\times {\bf S}^1_{\t{R}_5}$, but
 {\em with the monodromy }(\ref{pstep}) \cite{Witten}.

Remember the assumption $(p,q)=r(a,b)$ and $\mbox{gcd}(a,b)=1$.
Taking account of  the monodromy \eqn{pstep}, 
one can reformulate this Matrix theory (\ref{Zq-action})  
on the $b$-times covering circle ${\bf S}^1_{b\t{R}_5}$.
This ``screwing procedure'' is phrased  as follows:
For the set of functions $\{f_k(\th)\}_{k=0}^{q-1}$ with the relation
$f_k(\th+2\pi)=f_{k-p}(\th)$, we define 
\begin{eqnarray}
f_{\^{k}}(\th)=f_{k-pl}(\th-2\pi l) ~~~~&&2\pi l\leq \th \leq 2\pi (l+1)\nn
&&(l=0,\ldots,b-1~~~~\^{k}=0,\ldots,r-1) .
\end{eqnarray}
Then $f_{\^{k}}(\th)$ has 
the period $2\pi b$; $f_{\^{k}}(\th+2\pi b)=f_{\^{k}}(\th)$.
By this procedure, we can reduce the Matrix theory (\ref{Zq-action}) 
to an $A_{r-1}$ quiver theory on the ``long string'' 
${\bf S}^1_{b\t{R}_5}$; 
\begin{equation}
S={T_{D1}\o b}{1\o r}\sum_{\^{k}=0}^{r-1}
 \int dt\int_0^{2\pi b\t{R}_5}dx^1 \Tr_N L_{\^{k}} .
\label{long string}
\end{equation}

The fact that the theory has reduced to the $A_{r-1}$ quiver 
is not surprising, because we have already known that
$X_{p,q}$ actually possesses an  $A_{r-1}$-singularity.
The Higgs branch moduli space is known \cite{BHOO} 
to have the structure $Sym^N (ALE(A_{r-1}))$, which merely
corresponds to the picture that free D-particles 
move in the ALE space with $A_{r-1}$-singularity.

However, the Coulomb branch moduli, which describes  the 6-dimensional
dynamics decoupled from the bulk, is non-trivial and more interesting.

~

\subsection{Coulomb Branch Moduli Space and Fractional String}
\hspace*{4.5mm}
Recall that the Matrix theory \eqn{long string} is defined on  
${\bf R}  \times {\bf S}^1_{b\t{R}_5} \equiv 
{\bf R}  \times {\bf S}^1_{bl_s^2/R_5}$.  
The Coulomb branch $\M_V$ is hence parametrized 
by $\phi_m^i\in \left({\bf R}^4\times {\bf S}^1_{R_5/b}\right)^N$ 
(the Cartan components of 2-dimensional $N=(4,4)$ vectormultiplet and 
the Wilson line)\footnote
    {Precisely speaking,
we should define $\M_V$ as the quotient space 
by the isometry $\phi_m^i \riya \phi_m^i+\phi^{(0)}$.}.
Here the superscript $i(=0,\ldots,r-1)$ labels
the nodes of quiver, and $m(=1,\ldots,N)$ denotes the color index of $U(N)$. 

At tree level, the  metric  is diagonal:
 $\dsp ds^{2~(0)} ={TR_5\o qg_s^Al_s} \cob_{ij} \cob^{mn} \,
                            d\phi^i_m d\phi^j_n$.

Quantum correction for the metric of the Coulomb branch 
comes from the integration of the massive fields on this branch,
which are the off-diagonal
component of the vector multiplet $\vX^{i}_{mn}~(m\not =n)$ and 
the hypermultiplet $(Q^{i,i+1}_{mn},\t{Q}^{i+1,i}_{nm})$.
The masses  of these fields are 
given by 
$\phi^{ii}_{mn}$ and $\phi^{i,i+1}_{mn}$,
respectively, where we set $\phi^{ij}_{mn} \df \phi^i_m-\phi^j_n$. 

It is well-known that the correction exists only in the one-loop
level by SUSY cancellation (and in this case, we have no instanton correction).
Evaluating all the contributions of one-loop Feynman 
diagrams associated with the above massive modes,
we obtain the following result;  
\begin{eqnarray}
&&g^{mm~(1)}_{ii}=-2\sum_{n\not =m}G_1(\phi^{ii}_{mn}; bl_s^2/R_5)
 +\sum_{n,j\not =i}\hat{a}_{ij}G_1(\phi^{ij}_{mn}; bl_s^2/R_5) \nn
&&g^{mn~(1)}_{ii}=2G_1(\phi^{ii}_{mn}; bl_s^2/R_5) ~~~~(m\not =n) \nn
&&g^{mn~(1)}_{ij}=-\hat{a}_{ij}G_1(\phi^{ij}_{mn}; bl_s^2/R_5)~~~~(i\not = j)
\label{wittenC}
\end{eqnarray}
where the function
$G_1(\phi^{ii}_{mn}; bl_s^2/R_5)$ is written 
in terms of the modified Bessel function $K_1$;  
\begin{equation}
G_1(\phi,R)={1\o 2|\vX|^2}\left\{1+2\sum_{k=1}^{\infty}
m_k|\vX|K_1\bl(m_k|\vX|\br)\cos(m_ky)\right\} ,
\end{equation}
where $\phi=(\vX,y)\in {\bf R}^4\times {\bf S}^1_{R_5/b}$
(see  Appendix for the detail).
This is nothing but the Green function on 
${\bf R}^4\times {\bf S}^1_{R_5/b}$. 
$\hat{a}_{ij}$ stands for  the adjacency
matrix of the $A_{r-1}$ affine Dynkin diagram.
The above one-loop metric \eqn{wittenC} can be written in a simple form, 
\begin{eqnarray}
ds^{2~(1)} 
&=&{1\o 2}\sum_{m,n,i,j}a^{mn}_{ij}G_1(\phi^{ij}_{mn}; bl_s^2/R_5)
\bl(d\phi^{ij}_{mn}\br)^2
\label{wittenC2}
\end{eqnarray}
with $a_{ij}^{mn}=2\cob_{mn}\cob_{ij}-\hat{C}_{ij}$, where $\hat{C}_{ij}$
is the $A_{r-1}$ affine Cartan matrix.

The behavior of this metric is characterized by the Green function
$G_1(\phi ; bl_s^2/R_5)$.
It is worth remarking that it reflects the effects of the fractional 
string excitations.
In fact, let us give a naive estimation of metric and compare it
with above result.
Since $X_{p,q}$ is $\T^2$-fibered,  
if one consider the D0-probe theory
under the $X_{p,q}$-background, the loop calculation 
will need the summation of winding modes around this $\T^2$.
From the construction, this $\T^2$ may have  a non-trivial moduly.
But, in the limit that the TN-circle decompactifies 
$R_6 \rightarrow \infty$, this $\T^2$ becomes a simple rectangular
torus, and there survive only the winding modes $\sim nTR_5$.
On the other hand, we know $X_{p,q}$ has an $A_{r-1}$-singularity,
which will imply an $A_{r-1}$ quiver gauge theory on D0-brane.
The $A_{r-1}$ quiver theory with the summation of these winding
modes will give the Coulomb branch metric of the same form as
\eqn{wittenC}, {\em but with $G_1(\phi ; l_s^2/R_5)$ 
instead of $G_1(\phi ; bl_s^2/R_5)$.}  Of course, this is not
the correct result, since this naive estimation forgets the fractional 
windings $\dsp \sim n\frac{TR_5}{b}$ inside  the singular surface. 
The summation of these  modes will give the correct answer \eqn{wittenC}. 

Note the fact that the Witten's Matrix string becomes 
an  $A_{r-1}$ quiver  {\em after taking the screwing procedure}, and 
in this procedure,    
it automatically includes the ``maximally twisted sector''.
This  is no other than the excitations of fractional string. 

Although this consideration seems satisfactory, it feels somewhat
indirect for our purpose.
In the next section, we try to perform 
a more direct approach to the physics of fractional string,
that is, the Matrix string theory on the fractional string probe.

~


\section{Fractional String Probe}
\cleqn
\hspace*{4.5mm}
In this section, we shall try to construct the Matrix theory describing
the fractional string.
As we already mentioned, 
the fractional  string  is  realized as  the 
M2-brane wrapped around the fractional circle $\sim \e_{11}/b$
at the singularity of  
$({\bf S}^1\times {\bf C}^2)/\Z_q$. 
After the $X^{11}-X^5$ flip, the instanton string is represented by the
$D2$-brane wrapped around $\sim \e_{5}/b$. 
To obtain the desired theory, we consider the $D2$-brane
on $({\bf R}\times {\bf C}^2)/\Gamma$ whose world-volume 
is extended to $X^0,X^1,X^5$-direction.
We identify one spacial
coordinate $\si$ of the $D2$-brane world-volume and target space
coordinate $X^5$.
Thus, the theory on the
fractional string is given by the
$(2+1)$-dimensional $U(q)$ supersymmetric Yang-Mills theory projected
by the following $\Gamma$-action\footnote
      {In this section, we consider the single $D2$-brane probe
          ($U(1)$-gauge theory) to avoid unessential complexity.
        It is straightforward to include further the color 
        indices of $U(N)$, (which means we start from $U(qN)$-gauge 
        theory) but a little too cumbersome for our purpose.}.
The action of $\Gamma$ is given by 
\begin{eqnarray}
&&\al:~ \left\{
 \begin{array}{l}
 X^{\mu}_{I,J}\lf(\si-2\pi R_5 {a\o b}\ri)=\om^{I-J}X^{\mu}_{I,J}(\si)
  ,~~~(\mu=2,3,4) \\
 Q_{I,J}\lf(\si-2\pi R_5 {a\o b}\ri)=\om^{1+I-J}Q_{I,J}(\si) \\
 \t{Q}_{I,J}\lf(\si-2\pi R_5 {a\o b}\ri)=\om^{-1+I-J}\t{Q}_{I,J}(\si)
    \end{array}\right.  \\
&&\bt:~\left\{
 \begin{array}{l}
 X^{\mu}_{I,J}(\si+2\pi R_5)=X^{\mu}_{I,J}(\si),~~~(\mu=2,3,4) \\
 Q_{I,J}(\si+2\pi R_5)=Q_{I,J}(\si) \\
 \t{Q}_{I,J}(\si+2\pi R_5)=\t{Q}_{I,J}(\si)
    \end{array}\right.
\end{eqnarray}
where we suppressed the coordinates of the
$(2+1)$-dimensional world-volume  other than $\si$, and  $I,J$ run
from $0$ to $q-1$.

For the moment, we focus on the behavior of the field $Q_{I,J}$ under 
$\Gamma$-action.
By the projection corresponding to  the element $\bt^a\al^{b}$, $Q_{I,J}$ 
satisfies
\begin{equation}
Q_{I,J}(\si)=\big(\om^{b}\big)^{1+I-J}Q_{I,J}(\si).
\end{equation} 
From this relation, it follows that unless
\begin{equation}
J \equiv I+1 ~~({\rm mod}~r),
\end{equation}
matrix element $Q_{I,J}$ is zero.  In the following, we write
$I=i+mr$ where $i=0,\ldots, r-1$ and $m=0,\ldots,b-1$, and denote the
matrix element $Q_{I,J}$ as $Q_{m,n}^{i,j}$.

Next we make the projection by the element $\bt^{c}\al^{d}$ $(ad-bc=1)$:
\begin{eqnarray}
&&Q_{m,n}^{i,i+1}\lf(\si+2\pi {R_5\o b}\ri)=
 e^{-i\th_6(m-n)}Q_{m,n}^{i,i+1}(\si) ~,
~~~~~ (i=0,\ldots,r-2) \nn
&&Q_{m,n}^{r-1,0}\lf(\si+2\pi {R_5\o b}\ri)=
 e^{-i\th_6(m+1-n)}Q_{m,n}^{r-1,0}(\si)
\label{monoQ}
\end{eqnarray}
where we set 
\begin{equation}
\th_6=2\pi {d\o b} ,
\end{equation}
and this is no other than the $\theta$-angle of $(p,q)$ 5-brane 
gauge theory in the decoupling limit \cite{Witten,kol}.
We can easily find the boundary conditions of the other fields.
These  conditions for the non-vanishing fields are given by
\begin{eqnarray}
&&X^{i,i}_{m,n}\lf(\si+2\pi {R_5\o b}\ri)=
 e^{-i\th_6(m-n)}X^{i,i}_{m,n}(\si)~,
~~~(i=0,\ldots,r-1) \nn
&&\t{Q}^{i+1,i}_{m,n}\lf(\si+2\pi {R_5\o b}\ri)
 =e^{-i\th_6(m-n)}\t{Q}^{i+1,i}_{m,n}(\si)~,~~~(i=0,\ldots,r-2) \nn
&&\t{Q}^{0,r-1}_{m,n}\lf(\si+2\pi {R_5\o b}\ri)
 =e^{-i\th_6(m-n-1)}\t{Q}^{0,r-1}_{m,n}(\si).
\label{monoX}
\end{eqnarray}
Note that the $R_5/b$-periodicity of the modes which have no monodromies
simply means the fractionality of our Matrix string probe.
But the existence of monodromies leads to the peculiar behavior of
the fractional string. We will later discuss about this point.

One can understand the appearance of $\th_6$ in the monodromy relation
(\ref{monoQ}) (\ref{monoX}) by the following argument.  For this
purpose, we replace $X_{p,q}=({\bf S}^1\times{\bf C}^2)/\Z_q$ by $Y_{p,q}=({\bf
S}^1\times W)/\Z_q$ \cite{Witten} where $W$ is the charge one Taub-NUT
space (whose asymptotic behavior is described by the Hopf fibration).  
${\bf S}^1\times W$ is fibered over ${\bf R}^3$ with fibers
$\T^2[\e_5,\e_6]$, where we denote $\T^2[\e_5,\e_6]$ as the quotient of 
${\bf R}^2$ by the lattice generated by $\e_5=(2\pi R_5,0)$ and 
$\e_6=(0,2\pi R_6)$, and  ${\bf S}^1$
fiber of $W$ corresponds to $\e_6$.  

The fiber $E$ of $Y_{p,q}=({\bf S}^1\times W)/\Z_q$ is given by
$\T^2[\e_5,\e_6]/\Z_q$, where $\Z_q$ acts to the point 
${\bf x}=x^5\e_5+x^6\e_6$ on $\T^2[\e_5,\e_6]$ by ${\bf x} \riya
{\bf x}+(\e_6-p\e_5)/q$.  We can see that
$E=\T^2[\e_5,\t{\e}_6]$ with $\t{\e}_6=(\e_6-p\e_5)/q$.  The Taub-NUT
direction $\e_6$ is represented by the new basis $\{\e_5,\t{\e}_6\}$ as
$\e_6=p\e_5+q\t{\e}_6$.  We make $SL(2,\Z)$ transformation for the basis of
the lattice and take the Taub-NUT direction as one of the basis vector
of the lattice, which we denote $\va_6:=a\e_5+b\t{\e}_6 \equiv \e_6/r$. 
Then the fiber of $Y_{p,q}$ 
becomes  $E'=\T^2[a\e_5+b\t{\e}_6,c\e_5+d\t{\e}_6] \equiv
\T^2[\va_6,\va_5]$ (outside the singularity). 
Note that $\va_6$ and $\va_5$ correspond to
$\bt^a\al^{b},\bt^{c}\al^d \in \Gamma$, respectively.  
Under this
$SL(2,\Z)$ transformation, a $(p,q)$ 5-brane turns into a $D5$-brane in
the Type IIB theory picture.
The moduli parameter of $E$ and $E'$ is given by
\begin{eqnarray}
&&\tau(E)=-{a\o b}+i{R_6\o rbR_5} \, ,\\
&&\tau(E')={d\o b}+i{rR_5\o bR_6} \, .
\end{eqnarray}
The decoupling limit $\tau(E) \riya i\infty$ is realized by  
$R_6/R_5\riya \infty$, and in our situation this is always achieved
independent of a fixed value of $R_5$,  
since we want to take the ALE-limit $R_6 \rightarrow \infty$
   \footnote{In \cite{Witten}, the limit $R_5 \rightarrow 0$ 
              is rather called the ``decoupling limit''. 
          But in the stand point of this paper, it is more
          suitable that the ALE-limit $(R_6 \rightarrow \infty)$
          is  called so. In the limit $R_5 \rightarrow 0$,
           the bulk  gravity decouples very fast, 
          and hence we shall call it the ``strongly decoupling limit''.}.
This fact was first pointed out in \cite{LMS}. 
In this limit, $2\pi\tau(E')$ becomes the theta
angle $\th_6=2\pi d/b$ of the decoupled 6-dimensional theory.

In the $\e_5,\e_6$ basis, $\va_6,\va_5$ are expressed as  $\va_6=\e_6/r$ and
$\va_5=-\e_5/b+d\e_6/rb$. The monodromy (\ref{monoQ})(\ref{monoX}) can be 
understood from the relation
\begin{equation}
{1\o b}\e_5={\th_6\o 2\pi}\va_6-\va_5.
\label{slant}
\end{equation}
A fractional string is a membrane which is stretched between a
$\va_6$-side of the parallelogram corresponding to  $E'=\T^2[\va_6,\va_5]$
and the opposite $\va_6$-side, and extends along 
the direction  $\e_5/b$ which is perpendicular to $\va_6$. 
(In \cite{kol2}, this configuration of a membrane is called ``strip''.) 
From the relation (\ref{slant}), when one shifts $\si$ by $2\pi R_5/b$,
one does not come back to the same point on the torus but go to the point
shifted in the Taub-NUT direction. Because of the $\al$-projection,
the fields on a fractional string have the charge under the shift in 
the Taub-NUT direction. So the shift of $\si$ by $2\pi R_5/b$ leads to
the monodromy (\ref{monoQ}) (\ref{monoX}). This situation is
reminisecent of that in \cite{DH}, in which the noncommutative geometry
in Matrix theory\cite{noncom} is argued. So it may be interesting 
to discuss the relation between our fractional Matrix string 
and noncommutative geometry.

In summary, the field theory on the fractional string probe is
$(2+1)$-dimensional $U(b)$ $A_{r-1}$ quiver gauge theory on
${\bf R}^2\times {\bf S}^1_{R_5/b}$ with the monodromies 
(\ref{monoQ}) (\ref{monoX})\footnote
   {Of course, if we start from the $N$ D2 probes, we will obtain 
   $U(bN)$ $A_{r-1}$ quiver gauge theory with the similar monodromy as 
   \eqn{monoQ}, \eqn{monoX}.}.  

Note that  
$X^{i,i}$ and
$(Q^{i,i+1},\t{Q}^{i+1,i})~~(0\leq i\leq r-2)$ 
are periodic up to a gauge transformation (with monodromy)
, e.g. 
\begin{equation}
X^{i,i}(\si+2\pi R_5/b)=UX^{i,i}U^{-1}
\end{equation}
 where
$U={\rm diag}(1,e^{-i\th_6},\ldots,e^{-i(b-1)\th_6})\in U(b)$.
In other words, the monodromies of these fields can be absorbed 
into a suitable Wilson line.
However,
$(Q^{r-1,0},\t{Q}^{0,r-1})$ are {\em not\/} periodic up to the same gauge
transformation $U$. This implys that
 {\em all} the monodromies of this system {\em cannot\/} 
be replaced  with a VEV of Wilson line.
It is the essential feature of our
fractional Matrix string theory that there exist  monodromies
which can never  be  eliminated.

We will observe later  that thanks to this monodromy, 
the moduli space of our quiver gauge theory 
behaves in the expected manner for the physics of fractional string.

~

\subsection{Moduli Space of the Fractional String Theory}
\hspace*{4.5mm}
First, we consider the Higgs branch of the fractional Matrix string,
which represents the string moving in the ALE direction.
The vacuum condition is given by
\begin{equation}
Q^{i,i+1}\tQ^{i+1,i}-\tQ^{i,i-1}Q^{i-1,i}=\zeta_i {\bf 1}_{b\times b}
\label{vacb}
\end{equation}
We included the Fayet-Iliopoulos parameters $\zeta_i~(0\leq i\leq r-1)$
for each $U(1)$ subgroup of gauge group $U(b)_i~(0\leq i\leq r-1)$.  
These FI-parameters $\zeta_i$ should satisfy the relation
 $\sum_{i=0}^{r-1}\zeta_i=0$ for the consistency of (\ref{vacb}). 
For the quiver gauge theory with the monodromies
 (\ref{monoQ}) and (\ref{monoX}),
only the components $(Q_{n,n}^{i,i+1},\t{Q}^{i+1,i}_{n,n})~(0\leq i\leq r-2)$ 
and $(Q_{n,n+1}^{r-1,0},\t{Q}^{0,r-1}_{n+1,n})$ of the hypermultiplets
can have the zero-modes  and have the vacuum expectation values.
We define the gauge invariant variables $x$, $y$ and $z$ from 
these components;
\begin{eqnarray}
&&x=Q^{0,1}_{0,0}\tQ^{1,0}_{0,0} \nn
&&y=\prod_{n=0}^{b-1}\left(\prod_{i=0}^{r-2}Q_{n,n}^{i,i+1}\right)
 Q^{r-1,0}_{n,n+1} \nn
&&z=\prod_{n=0}^{b-1}\tQ^{0,r-1}_{n+1,n}\left(\prod_{i=0}^{r-2}
  \tQ^{i+1,i}_{n,n}\right)
\end{eqnarray}
By the relation (\ref{vacb}), other gauge invariant variables can be
expressed by $x$; 
\begin{eqnarray}
&&Q^{i,i+1}_{n,n}\tQ^{i+1,i}_{n,n}=x+a_i  \nn
&&Q^{r-1,0}_{n,n+1}\tQ_{n+1,n}^{0,r-1}=x+a_{r-1}
\end{eqnarray}
where $a_0=0$ and $a_i=\sum_{k=1}^{i}\zeta_i, ~(i=1,\ldots,r-1)$.
From these relations, the complex structure of
the Higgs branch is described by the equation:
\begin{equation}
yz=\left(\prod_{i=0}^{r-1}(x+a_i)\right)^b
\label{higgsb}
\end{equation}
This is nothing but the eqation of the $A_{q-1}$-ALE space.
However,  we should remark the following fact: In our quiver theory,
only the FI parameters {\em which are sufficient to resolve
the partial $\bz_r$-singularity \/} can be included. 
So, the $\bz_q/\bz_r \cong \bz_b$-singularity remaines at the 
every point in  the Higgs branch. 
This gives us the physical picture that only the $b$ joined 
fractional strings can freely move in the  ALE bulk space 
with the $A_{r-1}$-singularity.

It may be  meaningful to compare this result with that of the usual
quiver theory {\em without} monodromy.  
This is known \cite{BHOO} to have the structure
of a symmetric orbifold;
$Sym^b({\rm ALE}(A_{r-1}))$. This fact 
corresponds to the simple picture
that $b$ strings freely move in the $A_{r-1}$-ALE space,
and is not suited to  the behavior  of fractional string .

Next, let us  consider 
the Coulomb branch $\M_V^{\th_6}$ of the fractional string
theory. This branch should correspond to the fractional string moving inside
the $(p,q)$ 5-brane. 
As in the previous section, 
the Coulomb branch is parametrized by 
$\phi^i_m\in {\bf R}^3\times {\bf S}^1_{b\t{R}_5}$
$(i=0,\ldots,r-1,~m=0,\ldots,b-1)$ (three Cartan components 
of the vectormultiplet and one Wilson line around ${\bf S}^1_{R_5/b}$).

At the tree level, the metric of $\M^{\th_6}_V$ is diagonal:
$ds^{2~(0)} \propto \cob_{ij}\cob_{mn}d\phi^i_md\phi^j_n$.

We calculate the metric of $\M^{\th_6}_V$ to the one-loop order.
For the $U(q)$ $A_{r-1}$ quiver gauge theory without monodromy,
the metric of the Coulomb branch has the same form as (\ref{wittenC2})
with $G_1(\phi^{ij}_{mn},bl_s^2/R_5)$ replaced by 
$G_2(\phi^{ij}_{mn},R_5/b)$ defined as follows;
\begin{equation}
G_2(\phi;R)={T\o 4|\vX|}\left\{1+2\sum_{k=1}^{\infty}
e^{-m_k|\vX|}\cos(m_ky)\right\},
\end{equation}
where $\phi=(\vX,y)\in {\bf R}^3\times {\bf S}^1_{b\t{R}_5}$
 (see Appendix for the details). 
As in the previous case, this coincides with the Green function on 
${\bf R}^3\times {\bf S}^1_{b\t{R}_5}$, and was first introduced 
in \cite{diacone}.

Now we consider the theory with monodromy. 
The monodromy changes the mass of particles
which run around the loops.
Thus the one-loop metric of the fractional string theory
can be obtained by replacing $\phi^{ij}_{mn}$
by the following ``modified mass'' $\hat{\phi}^{ij}_{mn}$;
\begin{eqnarray}
&&\hat{\phi}^{ij}_{mn}=\hat{\phi}^i_m-\hat{\phi}^j_n,~~~~(i,j)\not =
(r-1,0),(0,r-1) \nn
&&\hat{\phi}^{r-1,0}_{mn}=-\hat{\phi}^{0,r-1}_{nm}=
\hat{\phi}^{r-1}_m-\hat{\phi}^0_n-(\vec{0},\th_6b\t{R_5})
\label{masshy}
\end{eqnarray} 
where $\hat{\phi}^i_m=\phi^i_m-(\vec{0},m\th_6b\t{R}_5)$.
The one-loop metric can be written as; 
\begin{equation}
ds^{2~(1)} ={1\o 2}\sum_{m,n,i,j}a_{ij}^{mn}
G_2(\hat{\phi}^{ij}_{mn};R_5/b)\bl(d\phi^{ij}_{mn}\br)^2 .
\end{equation}

Here we again 
emphasize that the effects of monodromies 
cannot be obtained by merely shifting the Wilson line, 
i.e.,  replacing $\phi^i_m$ by $\hat{\phi}^i_m$.
(This is due to the extra term for $\hat{\phi}^{r-1,0}_{mn}$ in 
the above expressions \eqn{masshy}.)
This statement corresponds to the fact that 
all fields in the fractional string theory cannot be made periodic 
simultaneously
by any gauge transformation.     

Now,
let us argue on the structure 
of singularity in the Coulomb branch $\M^{\th_6}_V$.
In general, the singularity of the moduli space is the point where the extra
massless particle appears. As is mentioned above, the mass of the
hypermultiplet is proportional to 
$\hat{\phi}^{i,i+1}_{m,m+1}$ (\ref{masshy}).

At the origin of the Coulomb branch $\phi^i_m=0$,
there are extra massless hypermultiplets 
$(Q^{i,i+1}_{m,m},\t{Q}^{i+1,i}_{m,m})~(0\leq i\leq r-2)$ and 
$(Q^{r-1,0}_{m,m+1},\t{Q}^{0,r-1}_{m+1,m})$ which have no monodromy.
Needless to say, the VEVs of these massless fields parametrizes
the Higgs branch above discussed.
That is, this branch emanates  from the origin of the Coulomb branch.     

However, there are other singularities in this branch.
One can immediately notice the existence of 
the next singular points $P_j$ $(j=0,\ldots,r-1)$ 
which are distributed $\bz_r$-symmetrically  around the origin;
\begin{equation}
P_j:~\left\{\begin{array}{ll}
  \phi^i_m=(\vec{0},m\th_6b\t{R}_5), &(i=0,\ldots,j) \\
  \phi^i_m=(\vec{0},(m+1)\th_6b\t{R}_5),&(i=j+1,\ldots,r-1)
       \end{array}\right.
\end{equation}
For example, $P_{r-1}$ is the point where $\hat{\phi}^i_m=0$. 
At $P_j$,    the hypermultiplet $(Q^{i,i+1},\t{Q}^{i+1,i})~(i\not =j)$ is
massless for every $U(b)$ index. (The number of massless particles is 
much larger than that of the origin!)
Nevertheless, the Higgs branch does {\em not \/} emanate from this
point.
In fact, all the components  of 
$(Q^{j,j+1},\t{Q}^{j+1,j})$ are massive, and so, we have
no non-trivial solution  for the equation of flat direction.
 
In the same way, we can find many other singular points in the Coulomb
branch. But, the point which can make a transition  to another branch 
is only the origin. Only at this point, the $b$ fractional strings can join
and generates the Higgs branch, that is, move into the ALE bulk.

For the end of this section, let us consider the asymptotic behavior of
the metric of the Coulomb branch in the limit $R_5\riya \infty$ and
$R_5\riya 0$. In the limit $R_5\riya \infty$, the world-volume of
$D2$-brane ${\bf R}^2\times {\bf S}^1_{R_5/b}$ becomes flat
three-dimensional space ${\bf R}^3$. In this limit, since
the effect of monodromy
vanishes, i.e. $\hat{\phi}^{ij}_{mn}\riya \phi^{ij}_{mn}$, the Coulomb
branch $\M_V^{\th_6}$ is the same as that of the 
ordinary $U(b)$ $A_{r-1}$ quiver gauge theory on ${\bf R}^3$ which
is known from the mirror symmery of the $d=3$ $N=4$ supersymmetric theory 
\cite{intri,HW}
to be the moduli space of $b$-instantons in  $SU(r)$ gauge theory.

On the other hand,
in the ``strongly decoupling limit'' 
$R_5\riya 0$, the mass of the field with
monodromy becomes infinite, so the excitations  of these fields decouple. 
The fields with no monodromies  do not decouple, and 
have the masses; 
$\hat{\phi}^{ij}_{mm}=\phi^{ij}_{mn}~(i,j)\not =(0,r-1),(r-1,0)$,
$\hat{\phi}^{r-1,0}_{m,m+1}=\phi^{r-1,0}_{m,m+1}$.   
The metric of the Coulomb branch is then reduced to
\begin{eqnarray}
ds^{2~(0)} &=&{1\o 2}\sum_m
\sum_{{\scriptstyle (i,j)\ne (r-1,0)}\atop{\scriptstyle ~~~~~~(0,r-1)}}
\hat{a}_{ij}G_2(\phi^{ij}_{mm},R_5/b)\bl(d\phi^{ij}_{mm}\br)^2
+\sum_mG_2(\phi^{r-1,0}_{m,m+1},R_5/b)\bl(d\phi^{r-1,0}_{m,m+1}\br)^2 \nn
&=& {1\o 2}\sum_{I,J=0}^{q-1}\hat{a}_{IJ}
    G_2(\phi_{IJ},R_5/b)\bl(d\phi_{IJ}\br)^2
\end{eqnarray}
where $\phi_I=\phi^i_m$ with $I=i+mr$ and $\hat{a}_{IJ}$ is the adjacency
matrix of the $A_{q-1}$ affine Dynkin diagram.
This metric is the same as that of the $U(1)$ $A_{q-1}$ quiver gauge
theory on ${\bf R}^2$.

One can understand this phenomenon both from
the M-theory and Type IIB pictures.
In the M-theory picture, $({\bf S}^1_{R_5}\times {\bf C}^2)/\Z_q$
becomes  ${\bf C}^2/\Z_q$ in the decoupling limit, and so the theory on the 
fractional string reduces to the usual $A_{q-1}$ quiver theory.
In the Type IIB picture, because the tension of the NS5-brane is much larger
than that of the $D5$-brane in this  limit, a $(p,q)$-fivebrane
becomes effectively $q$ NS5-branes, of which T-dual picture 
is of course the $A_{q-1}$ ALE.

~

\section{Discussion}
\hspace*{4.5mm} 
We have studied two Matrix string theories as the probe of
the $(p,q)$ 5-brane: Witten's Matrix string theory and the fractional
string theory. These theories are respectively
$(1+1)$-dimensional $U(N)$ $A_{q-1}$
quiver gauge theory and $(2+1)$-dimensional $U(bN)$ $A_{r-1}$ quiver
gauge theory {\em with monodromy\/}. 

In the Witten's Matrix string theory the monodromy acts as a diagram
automorphism (clock-shift) of the extended Dynkin diagram for quiver,
which reduces the theory to the $A_{r-1}$ quiver.
In this screwing procedure, it naturally 
incorporate the excitations of fractional string. 

On the other hand,
the monodromies in the fractional Matrix string theory have different 
forms  - the phase shifts of vector and hypermultiplets, which is 
discussed in \cite{BCD}.
They play an essential role in the fractional Matrix string. 
Thanks to them, we  can  realize the peculiar  behavior of fractional 
string from the viewpoint of Matrix theory. Our analyses of 
moduli spaces confirm  the following expectation;
a single  fractional string cannot move  away from the singular surface
(the world-brane of $(p,q)$ 5-brane), and  
only the joined fractional strings which have the equal tension  
to a fundamental string can do.

The quiver Matrix theories are ``magnetic'' 
(in the usual convention of terminology) formulations for
the brane theory probing  5-branes.  
Thus, it may be an  interesting 
task to construct the fractional Matrix string 
theory as the ``electric'' theory. In this framework,
the decoupled 6-dimensional physics should be  described 
as Higgs branch, which is the instanton moduli space
(this is a tautology, since the fractional string is an instanton from
the beginning!), and the joined fractional strings moving away 
from the $(p,q)$ 5-branes corresponds to the Coulomb branch.

As we mentioned in section 1, this is not an easy problem,
since the instanton string is not a D-brane in general.
Moreover, in our discussion, so-called 
the Mirror symmetry for 3-dimensional 
gauge theory \cite{intri,BHOO}
cannot be applicable in the exact sense.
This is because, in the case when $R_5$ is finite, 
the quantum moduli space of Coulomb branch has no continuous isometry,
and so, the electric-magnetic duality is not reduced to the simple 
``T-dual'' transformation \cite{diacone}.
(The case $R_5 = \infty$, the monodromy lose its meaning and 
our fractional Matrix string reduces to the usual D2-probe.)

Although difficulties exist,
we believe it meaningful to construct the electric theory
on account of  a few reasons.
First, for the 5-branes with a {\em irrational\/} $\theta$-angles,
the ALE description failes. 
Nevertheless, the decoupled 6-dimensional theory 
can be similarly defined in the 5-brane framework \cite{kol}.
This implies that  the electric formulation of ``instanton Matrix string''
(which can have an irrational tension in general) 
may be also applicable for the irrational cases.
Second, let us note  the following fact:  The configurations of 
many D5s with non-vanishing $\theta$-angle and many  instanton  strings 
are natural generalizations of those corresponding to
the Mardacena's $AdS_3 \times S^3$ SCFT \cite{ads3}. We emphasize that 
in the cases of non-zero $\theta$-angle, $AdS_3$ CFT 
does {\em not\/} correspond to the system of D5$+$D1, {\em but\/} to  
D5$+$instanton strings.
In this meaning, the electric formulation
of instanton Matrix string  may add a  new perspective
to the study of $AdS_3$ CFT.

~

\centerline{{\bf Acknowledgements}}
The work of ~K. O.~ is supported in part by JSPS Research Fellowships for Young
Scientists.

~

~

~

\appendix
\noindent
{\Large {\bf Appendix}}
\section{One-loop integral on ${\bf R}^n\times {\bf S}^1_R$}
\hspace*{4.5mm}
One vector multiplet of $(n+1)$-dimensional supersymmetric gauge theory
 with 8 supercharges contains $(5-n)$ scalar fields, which we denote
 $\vX\in {\bf R}^{5-n}$.  The metric of the Coulomb branch is written in
 terms of the one-loop integral on ${\bf R}^n\times {\bf S}^1_R$,
\begin{equation}
G_n(\phi;R)={1\o R}\sum_{k=-\infty}^{\infty}\int {d^np\o (2\pi)^n}
T^2\left[p^2+{1\o R^2}(k+TRy)^2+\bl(T|\vX|\br)^2\right]^{-2}
\end{equation}
where $\dsp y=T^{-1}\int_{{\bf S}^1_R}{A\o 2\pi R}$ 
is the Wilson line, which has
periodicity $y\sim y+2\pi\t{R}$ with $\t{R}=1/2\pi RT=l_s^2/R$, and
$\phi=(\vX,y)$ is the coordinate of ${\bf R}^{5-n}\times 
{\bf S}^1_{\t{R}}$. 
After the Poisson resummation, $G_n(\phi;R)$ is rewritten as
\begin{eqnarray}
G_n(\phi;R)&=&{1\o 2}\left({2\pi\o T^2}\right)^{\nu-1}
\sum_{k=-\infty}^{\infty}
\left({|m_k|\o |\vX|}\right)^{\nu}K_{\nu}(|m_k\vX|)e^{im_ky}  \nn
&=&{1\o 2}\left({2\pi\o T^2}\right)^{\nu-1}
\left\{{2^{\nu-1}\Ga(\nu)\o |\vX|^{2\nu}} + 
2\sum_{k=1}^{\infty}\left({m_k\o |\vX|}\right)^{\nu}K_{\nu}(m_k|\vX|)
\cos(m_ky)\right\}.
\end{eqnarray}
Here we defined  $m_k=2\pi RTk=k/\t{R}$ and 
$\nu=(3-n)/2$. $K_{\nu}(z)$ is the modified Bessel function;
\begin{equation}
K_{\nu}(z)={1\o 2}\left({z\o 2}\right)^{\nu}\int_0^{\infty}\!dt~ t^{-\nu-1}
\exp\left(-t-{z^2\o 4t}\right).
\end{equation}
Note that $G_n(\phi;R)$
is the Green function on ${\bf R}^{5-n}\times {\bf S}^1_{\t{R}}$;
\begin{equation}
\bl(\Delta_{\vX}+\del_y^2\br)G_n(\phi;R) 
= -{T\o 2R}\left({2\pi\o T}\right)^{2\nu}
\cob(\vX)\cob(y)
\end{equation}

In the limit $R\riya \infty$, $G_n(\phi;R)$ behaves like
\begin{equation}
G_n(\phi;R) \sim {1\o 2}\left({4\pi\o T^2}\right)^{{1-n\o 2}}
{\Ga\left({3-n\o 2}\right)\o |\vX|^{3-n}}
\end{equation}
and in the limit $R\riya 0$,
\begin{equation}
2\pi RG_n(\phi;R) \sim {1\o 2}\left({4\pi\o T^2}\right)^{{2-n\o 2}}
{\Ga\bl({4-n\o 2}\br)\o |\phi|^{4-n}}
\end{equation} 

In section 2, we need $G_1(\phi,R)$ which is given by
\begin{equation}
G_1(\phi,R)={1\o 2|\vX|^2}\left\{1+2\sum_{k=1}^{\infty}
m_k|\vX|K_1\bl(m_k|\vX|\br)\cos(m_ky)\right\}
\label{g1}
\end{equation}
and $G_2(\phi,R)$ is relevant for the fractional string theory, 
\begin{equation}
G_2(\phi;R)={T\o 4|\vX|}\left\{1+2\sum_{k=1}^{\infty}
e^{-m_k|\vX|}\cos(m_ky)\right\}
\label{g2}
\end{equation}
where we used the fact that $\dsp K_{1/2}(z)=\sqrt{{\pi\o 2z}}e^{-z}$.

\newpage


\baselineskip 6mm
\begin{thebibliography}{99}
\bibitem{DVV}
R. Dijkgraaf, E. Verlinde and H. Verlinde,
``Matrix String Theory'', Nucl. Phys. {\bf B500} (1997) 43, hep-th/9703030. 

\bibitem{NS5A}
E. Witten, ``On the Conformal Field Theory of the Higgs Branch'',
JHEP {\bf 07} (1997) 003, hep-th/9707093; \\
O. Aharony, M. Berkooz, S. Kachru, N. Seiberg, E. Silverstein, ``Matrix
 Description of Interacting Theories in Six Dimensions'', 
Adv. Theor. Math. Phys. {\bf 1} (1998) 148, hep-th/9707079; \\
R. Dijkgraaf, E. Verlinde, H. Verlinde,
``5D Black Holes and Matrix Strings'',  
Nucl. Phys. {\bf B506} (1997) 121, hep-th/9704018. 

\bibitem{diacone}
D-E. Diaconescu and N. Seiberg,
``The Coulomb Branch of (4,4) Supersymmetric Field Theories in Two 
Dimensions'', JHEP {\bf 07} (1997) 001, hep-th/9707154.


\bibitem{new-th}
N. Seiberg, ``New Theories in Six Dimensions and Matrix Description 
of M-theory on $T^5$ and $T^5/Z_2$'', Phys. Lett. {\bf B408} (1997) 98,
hep-th/9705221.

\bibitem{kol2}
O. Aharony, A. Hanany and B. Kol,
``Webs of (p,q) 5-branes, Five Dimensional Field Theories and
Grid Diagrams'', JHEP {\bf 01} (1998) 002, hep-th/9710116.

\bibitem{Witten}
E. Witten, ``New ``Gauge'' Theories in Six Dimensions'',
JHEP {\bf 01} (1998) 001, hep-th/9710065.

\bibitem{kol}
B. Kol, ``On 6d ``Gauge'' Theories with Irrational Theta Angle'',
hep-th/9711017.

\bibitem{BFSS}
T. Banks, W. Fischler, S. H. Shenker, and L. Susskind,
``M Theory As A Matrix Model: A Conjecture'',
Phys. Rev. {\bf D55} (1997) 5112, hep-th/9610043.

\bibitem{d-moore}
M. R. Douglas and G. Moore, ``D-branes, Quivers, and ALE Instantons'',
hep-th/9603167.

\bibitem{douglas}
M. R. Douglas, ``Enhanced Gauge Symmetry  In M(atrix) Theory'',
JHEP {\bf 07} (1997) 004, hep-th/9612126; \\
D-E. Diaconescu, M. R. Douglas and J. Gomis,
``Fractional Branes and Wrapped Branes'',
JHEP {\bf 02} (1998) 013, hep-th/9712230.



\bibitem{taylor}
W. Taylor,
``D-brane field theory on compact spaces'',
Phys. Lett. {\bf B394} (1997) 283, hep-th/9611042.

\bibitem{LMS}
A. Losev, G. Moore and S.L. Shatashvili,
``M \& m's'', hep-th/9707250

\bibitem{DH}
M. R. Douglas and C. Hull,
``D-Branes and the Noncommutative Torus'', JHEP {\bf 02} (1998) 008,
 hep-th/9711165.

\bibitem{noncom}
A. Connes, M. R. Douglas and A. Schwarz,
``Noncommutative Geometry and Matrix Theory: 
Compactification on Tori'',
JHEP {\bf 02} (1998) 003, hep-th/9711162; \\
P.-M. Ho, Y.-Y. Wu and Y.-S. Wu,
``Towards a Noncommutative Geometric Approach to Matrix
Compactification'',
hep-th/9712201; \\
Y.-K. E. Cheung and M. Krogh,
``Noncommutative Geometry from 0-branes in a Background B-field'',
hep-th/9803031; \\
T. Kawano and K. Okuyama,
``Matrix Theory on Noncommutative Torus'',
hep-th/9803044.

\bibitem{BHOO}
J. de Boer, K. Hori, H. Ooguri and Y. Oz,
``Mirror Symmetry in Three-Dimensional Gauge Theories,
Quivers and D-branes'', hep-th/9611063.

\bibitem{intri}
K. Intriligator and N. Seiberg,
``Mirror Symmetry in Three Dimensional Gauge Theories'',
Phys. Lett. {\bf B387} (1996) 513, hep-th/9607207.


\bibitem{HW}
A. Hanany and E. Witten,
``Type IIB Superstrings, BPS Monopoles, And Three-Dimensional Gauge
 Dynamics'', Nucl. Phys. {\bf B492} (1997) 152, hep-th/9611230. 

\bibitem{BCD}
D. Berenstein, R. Corrado and J. Distler,
``Aspects of ALE Matrix Models and Twisted Matrix Strings'',
hep-th/9712049.



\bibitem{ads3}
J. Mardacena, 
``The Large N Limit of Superconformal field theories and
supergravity'', 
hep-th/9711200; \\
J. Mardacena and A. Strominger,
``$AdS_3$ Black Holes and a Stringy Exclusion Principle'' 
hep-th/9804085; \\
E. J. Martinec,
``Matrix Models of AdS Gravity'',
hep-th/9804111.



\end{thebibliography}
\end{document}